# Confinement Loss in Hollow-Core Negative Curvature Fiber: a Multi-Layered Model

**Yingying Wang[1] and Wei Ding[2,*]**

[1]*Beijing Engineering Research Center of Laser Technology, Institute of Laser Engineering, Beijing University of Technology*，*Beijing 100124, China*

[2]*Laboratory of Optical Physics, Institute of Physics, Chinese Academy of Sciences, Beijing 100190, China*

*[*wding@iphy.ac.cn](*wding@iphy.ac.cn)*

**Abstract:** Simple structures are always a pursuit but sometimes not easily attainable. It took researchers nearly two decades for conceiving the structure of single-ring hollow-core negative curvature fiber (NCF). Recently NCF eventually approaches to the centre of intense study in fiber optics. The reason behind this slow-pace development is, undoubtfully, its inexplicit guidance mechanism. This paper aims at gaining a clear physical insight into the optical guidance mechanism in NCF. To clarify the origins of light confinement, we boldly disassemble the NCF structure into several layers and develop a multi-layered model. Four physical origins, namely single-path Fresnel transmission through cascaded interfaces, near-grazing incidence, multi-path interference caused by Fresnel reflection, and glass wall shape are revealed and their individual contributions are quantified for the first time. Such an elegant model could not only elucidate the optical guidance in existing types of hollow-core fibers but also assist in design of novel structure for new functions.

## 1.  Introduction

After 20 years development of hollow-core photonic crystal fiber (HC-PCF) [1,2], a type of fiber structure as simple as a ring of 6-8 untouched thin tubes surrounding the hollow core [3-7] has very recently emerged as the cutting-edge research area in fiber optics. This type of hollow-core fiber (HCF), referred to as hollow-core negative curvature fiber (NCF) in this paper (or more broadly named hollow-core anti-resonant fiber (ARF)) shows similar level of transmission loss and single modeness with the maturely developed hollow-core photonic-bandgap fiber (PBGF) and outperforms the PBGF in terms of broadband light guidance and high laser damage threshold. It has found plenty of interdisciplinary applications in areas ranging from ultra-intense pulse delivery [8,9], single-cycle pulse generation [10,11], low latency optical communication [5], UV light sources [12,13], mid-IR gas lasers [14] to biochemical sensing [15,16], quantum optics [17] and mid-IR to Terahertz waveguides [18,19]. In some of these applications, NCF has the potential to revolutionize the research field by realizing unprecedented performances, e.g. ultralow transmission loss [20] for optical fiber communication.

Compared with NCF, PBGF possesses a much more complicated cladding structure consisting of 5-8 layers of triangular lattices [1]. While, its guidance mechanism has been fully elucidated [21]. In the periodical cladding region, Bloch waves replace plane waves to play the role of energy carrier, and the density of radiative states (DOS) can be modified or engineered via periodic structures. By opening out-of-plane photonic bandgaps (PBG) beneath the light line and exploiting longitudinal wavevector matching condition, all the outwardly-propagating passageways from the air core can be blocked. This unambiguous explanation of the guidance mechanism opens the door for PBGF design and has led to the rapid development of PBGF in the first decade of 21[st] century. Nowadays, it has been verified in both experiment and theory that the state-of-art attenuation level of PBGF (1.2dB/km at 1620nm [22]) is not limited by confinement loss (CL), but mainly by surface scattering loss.

The development of ARF is at a much slower pace because of the inexplicit guidance mechanism. When Kagome fiber (a type of ARF) was introduced in 2002 [23], people took it for granted that this fiber was lossier than PBGF, since its DOS diagram shows no stop band but several low-DOS spectral regions [24]. In 2010, a breakthrough in the attenuation figure of a delicately designed Kagome fiber unexpectedly triggered a boost of research on ARF [25]. By changing the core-surround shape from circle to hypocycloid [26], the attenuation figure of Kagome fiber was surprisingly decreased from 1000dB/km to 40dB/km [27] at 1550nm and recently to 12.3dB/km at 1064nm [28]. Inspired by this finding, more designs and fabrications on hypocycloidal or negative curved core shape ARFs (denoted as NCF) have proliferated in the second decade of the 21[st] century [3-7, 20, 29-33]. Nowadays, in most ARFs, the translational periodic lattice structure has been replaced by arrangement of untouched tubes with generally cylindrical symmetry. The attenuation level of NCF has reduced to 7.7dB/km at 750nm [3], nearly one order of magnitude lower than reported PBGFs guiding below 800nm. Despite these achievements, how such a simple NCF structure can attain a low CL remains an intriguing open question.

Considerable theoretical efforts have been devoted to study the optical effects occurring in ARF, such as capillary/Bragg fiber theory [34,35], anti-resonant reflecting optical waveguide (ARROW) model [36], DOS calculation [3], inhibited coupling analogy [37], tube lattice fiber model [38] and concentric rings model [39-43]. These models are able to predict the high and low loss spectral regions, the dispersion curves, the conditions for single modeness, and the bending loss performances [44]. However, when referring to the level of CL, all of these models fail to provide a quantitative interpretation. Precise CL information severely relies on numerical method with little physical insight. The lack of a quantitative model for ARF has substantially impeded the search for experimentally realizable better-performance structures. Up to now, the fundamental CL limit of ARF has still not been elucidated.

The purpose of this paper is to develop an intuitive model to gain physical insights and quantify different origins for CL in ARFs, specifically in NCF. The other loss origins, e.g. scattering, absorption, and bending, will be ignored. Our analysis and calculation focus on the fundamental ($HE_{11}$) core mode. We deal with annular fibers and NCFs in parallel and disassemble their structures into a series of layers in the radial direction. In this process we give out analytic expressions and numerical calculations to validate our multi-layered model. Four light confinement origins are identified: single-path Fresnel transmission, near-grazing incidence, multi-path interference caused by Fresnel reflection, and glass wall shape. The significance of each factor is explicitly presented, thanks to this multi-layered model. Although the basic principles of optical leaky waves are well known [45, 46], quantitative analysis of CL has only been successful in the Bragg fiber (referred to as annular fiber in this paper) [42, 43]. The failure of the same attempts in Kagome and tube lattice ARFs may be partly attributed to severe Fano resonances and partly to the adoption of translational symmetry [24,37,38]. To our knowledge, this is the first time that the composite light confinement effects in a NCF are combed and quantitatively analyzed. In this paper, we concentrate on the structure of single ring of eight tubes. Nevertheless, the method is probably universal and has the potential to extend to other valuable ARFs, such as, ice-cream shaped fiber, Kagome fiber, nested fiber, etc.

The paper is organized as follows. In section 2, we revisit the multi-layered annular fiber and identify different factors that influence the light confinement. We hypothetically separate the Fresnel transmission and reflection processes, paving a simple and straightforward way for quantifying their individual effects. In section 3, we turn our attention to the specificity of the NCF. We boldly disassemble its structure into a series of layers. The layer-by-layer comparison between the annular fiber and the NCF undoubtfully visualizes a new engineering paradigm, the shape of the glass wall. In the last section, we draw a conclusion and give a prospect to the employment of our model.

## 2. Light confinements in multi-layered annular fiber

Light confinement in a dielectric fiber can be realized by two types of reflection at interfaces: Fresnel reflection and total internal reflection (TIR). These interfaces hamper the communication between both sides. In general, TIR induced light confinement (via guided modes) is more efficient since all the passageways for light transmission are blocked. For Fresnel reflection induced light confinement (via leaky modes), despite of some inter-medium interfaces being poor reflectors at normal incidence, at near-grazing incidence however, it also becomes highly efficient. The silica/air interface with the refractive index $n = 1.45/1$ is such a system. Every time optical ray impinges on such an interface, it bifurcates, part of it being transmitted and part of it being reflected. The transmitted ray ($T$) delivers a fraction of incidence energy ($I$) to the other side and forms the dissipation as the light propagates in the z direction. If we hypothetically neglect the consequence of multiple (Fresnel) reflections, the single-path light transmission from the core region to the outside can be expressed as an exponential function of the number of interfaces, as illustrated in Fig. 1. The scaling factor can be approximately given by

$$\begin{cases} \dfrac{T_s}{I} = \dfrac{T_s{'}}{T_s} = \dfrac{T_s{''}}{T_s{'}} \approx 4\sin\theta_z \cdot \dfrac{1}{\sqrt{n^2-1}} & (s-Pol.) \\ \dfrac{T_p}{I} = \dfrac{T_p{'}}{T_p} = \dfrac{T_p{''}}{T_p{'}} \approx 4\sin\theta_z \cdot \left( \dfrac{1}{\sqrt{n^2-1}} + \sqrt{n^2-1} \right) & (p-Pol.) \end{cases}. \quad (1)$$

Here, we adopt a slab geometry at two polarization states, and the glancing angle in the air layer ($\theta_z$) is treated as a small quantity (see the derivation in Appendix A). For a cylindrical fiber, to the first-order approximation, we only consider meridional rays. A hybrid core mode, such as the fundamental $HE_{11}$ mode, can be decomposed in equal parts to $s$- and $p$-polarized waves with

$$\dfrac{\overline{T}}{I} = \dfrac{T_s + T_p}{2I} \approx \dfrac{(4\sin\theta_z)^N}{2} \left\{ \left( \dfrac{1}{\sqrt{n^2-1}} \right)^N + \left( \dfrac{n^2}{\sqrt{n^2-1}} \right)^N \right\}. \quad (2)$$

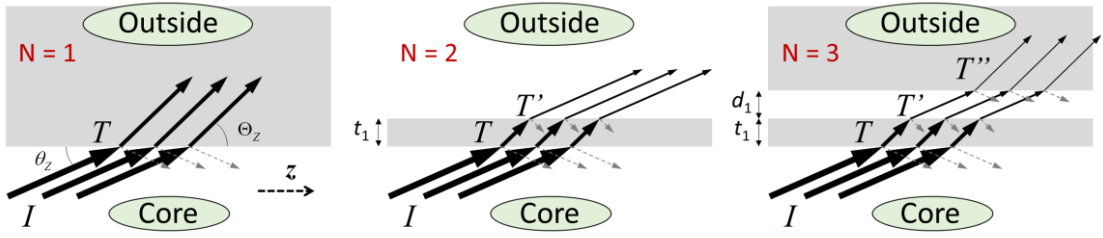

**Fig. 1**. A schematic drawing of multi-layered slab structures showing how the transmitted ray from the core to the outside is exponentially decayed. The outermost transmitted ray forms the attenuation of the core mode as it propagates in the $z$ direction. $N$ is the number of interfaces. $\theta_z$ and $\Theta_z$ stand for the glancing angles in the air and the glass layers respectively.

In order to achieve strong light confinement, the optical ray leaked to the outermost medium should be suppressed to be as little as possible. From Eq. (2), two conditions are favored: 1) a large number of interfaces, $N$; and 2) a small glancing angle in the air layer, $\theta_z$. These two conditions have actually been fulfilled in ARFs. For example, in comparison with conventional capillary, most ARFs encompass more inter-medium interfaces in the radial direction. On the other hand, compared with PBGFs, ARFs usually work in the large core size regime to ensure a small glancing angle. It has been demonstrated in many reported scaling laws [34,42,47] that the CLs of ARFs decrease with the ratio of the core radius ($a$) to the vacuum wavelength ($\lambda$), which is inversely proportional to $\sin\theta_z$.

In above treatment, Fresnel reflection and consequent multi-path interference have been neglected. Taking them back into account, these effects lead to the 3$^{rd}$ favored condition for strong light confinement -- the famous ARROW (anti-resonant reflecting optical waveguide) criterion [36]. Note that the parameters of layer thicknesses, i.e., $t_1$, $d_1$, in Fig. 1, influence this condition ONLY. In order to solely evaluate the effect of ARROW, we define a figure-of-merit (FOM) to normalize the attenuation coefficients (in unit of dB/m) of a multi-layered annular fiber ($\alpha_{fiber}$) with that of a capillary ($\alpha_{capillary}$), both having the same core diameters

$$FOM(dB) = 10 \cdot Log_{10}\left\{\frac{\alpha_{fiber}}{\alpha_{capillary}} \cdot \left(\frac{4\sin\theta_z}{\sqrt{n^2-1}}\right)^{-N+1} \cdot \frac{1+n^2}{1+n^{2N}}\right\}. \qquad (3)$$

Here, all the attenuation coefficients $\alpha$'s are calculated by using the transfer matrix method [35], and $\sin\theta_z \approx \sqrt{1-\text{Re}(n_{eff})^2}$ is derived from the complex effective index $n_{eff}$. The influence of the modal index, i.e., that of the glancing angle, is naturally excluded from this FOM, and via this formula we can individually analyse the effects of Fresnel transmission and reflection combined with Eq. (2). Furthermore, we introduce two ansatz variables of $U_{glass} = 2t_i\sqrt{n^2-\text{Re}(n_{eff})^2}/\lambda$ and $U_{air} = 2d_i\sqrt{1-\text{Re}(n_{eff})^2}/\lambda$ to construct the expression of $FOM(U_{glass}, U_{air})$. Different with the popularly used normalized frequency $F = 2t_i\sqrt{n^2-1}/\lambda$, these two ansatz variables characterize the transverse normalized frequencies in glass and air layers respectively. In calculating Eq. (3), we fix $t_i$ and $d_i$ to be 0.24μm and 10μm respectively and continuously alter the wavelength (λ) and the core radius (*a*). As a result, Re($n_{eff}$) varies and $U_{glass}$ ($U_{air}$) is adjusted across different ARROW bands (see Fig. 2). We note that, for an annular fiber, $U_{air}$ is nearly insensitive with the wavelength because of Re($n_{eff}$) $\approx \sqrt{1-(u_{01}\lambda/2\pi a)^2}$ (so that $U_{air} \approx d_i u_{01}/\pi a$) with $u_{01}$ the first null of the zeroth-order Bessel function [34].

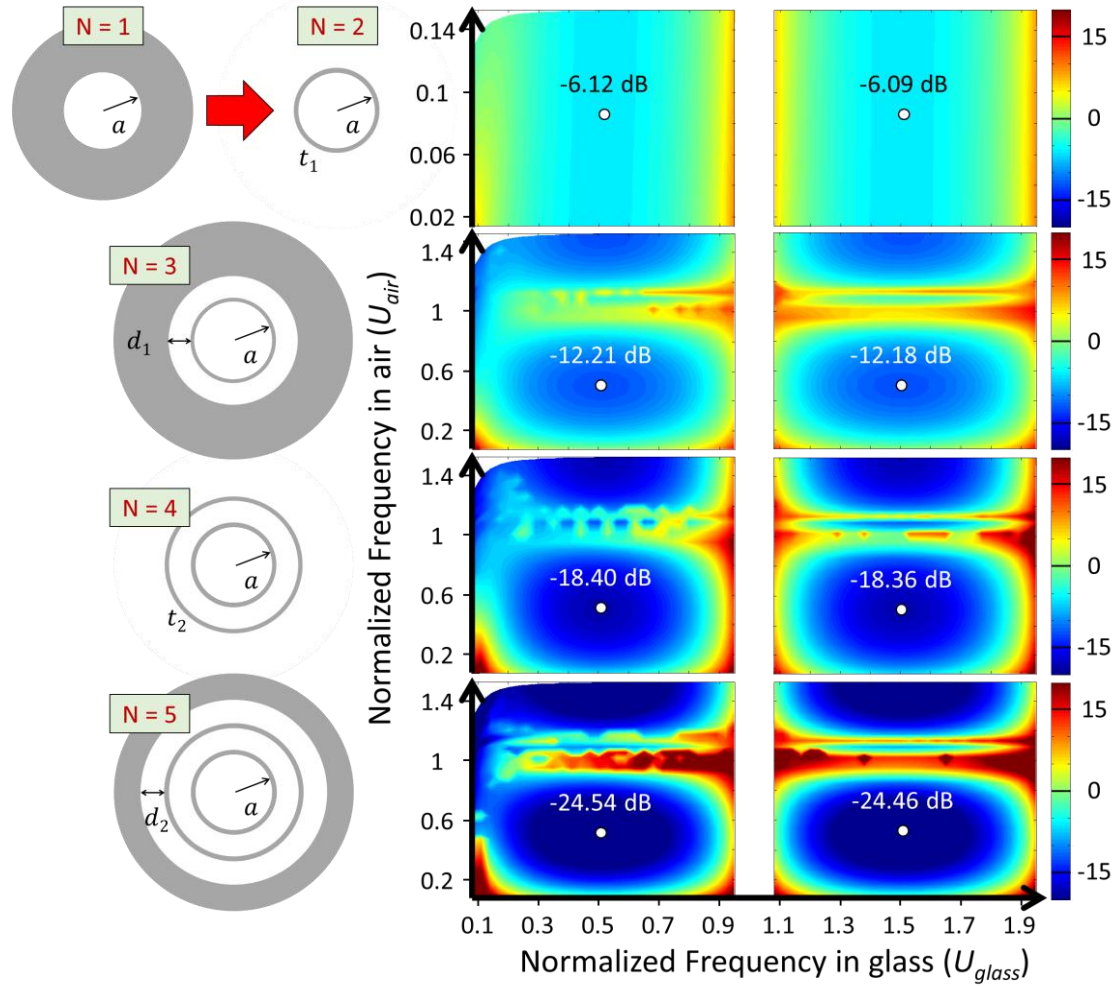

**Fig. 2**. Figure-of-merit (*FOM*) of ARROW effect as a function of transverse normalized frequencies $U_{glass}$ and $U_{air}$. In the first row the Y-axis is $2\sqrt{1-\text{Re}(n_{eff})^2}/\lambda$. From the top to the bottom row, the maximum FOM is sequentially increased by roughly 6dB for each anti-resonant layer.

In Fig. 2 we see the ARROW effect appears when N ⩾ 2, and the contributions from different layers, whether the glass or the air layers, can be accumulated. While the near-resonance regions (in the vicinity of $U_{glass}$ ($U_{air}$) = 1,2…) are detrimental for light confinement, in the anti-resonant regions, the attenuation coefficients are reduced by virtue of the ARROW effect. However, very interestingly, the maximum benefit from one anti-resonant layer is only ~6dB. This value is originated from the factor 4 in Eqs. (1-3), and seems not relevant with the order of ARROW band and the dielectric material. Detailed derivation is given in Appendix B.

## 3. Light confinements in negative curvature fiber

In the last section, we decompose and quantify (formulate) different origins in the light guiding process. In an annular fiber, three effects (single-path Fresnel transmission through cascaded interfaces, near-grazing incidence, and multi-path interference caused by Fresnel reflection) are identified. For an arbitrarily-shaped ARF, if one considers the radial radiation from the core to the outside, it is in many aspects in common with the annular fiber. However, there appears a new engineering paradigm, i.e. the shape of glass wall, which could also influence light confinement. In this section, we take the single ring nodeless NCF as an example and inspect its CL in analogy with the annular fiber. We disassemble the NCF structure layer by layer in the radial direction. Every glass tube is split into one negatively-curved half tube and one positively-curved half tube to form two glass layers, and all the air holes enclosed by glass tubes are viewed as an entire air layer even though they are not connected with each other.

As mentioned in the last section, the effective index of the core $HE_{11}$ mode is relevant to the glancing angle $\theta_z$. To gain a fair comparison of CL, we firstly design appropriate core diameters to ensure the $Re(n_{eff})$ of different fibers are equivalent. In Fig. 3 the diameters of the annular fiber and the NCF are $2a = 30\mu m$ and $2a' = 28.66\mu m$ respectively, guaranteeing the coincidence of their (leaky) modal indices ($\Delta Re(n_{eff}) < 2\times 10^{-6}$ in the wavelength region 520nm - 1500nm). Secondly, we keep the air layers of the two fibers in nearly optimum ARROW condition, $U_{air} \approx 0.5$. For the annular fiber this requires an air layer thickness $d \approx 10\mu m$ ( $d_{opt} \approx \lambda / \left(4\sqrt{1-n_{eff}^2}\right) \approx \frac{\pi}{2u_{01}}a$ [47]), and for the NCF the air hole inner diameter is set to $d' \approx 16\mu m$. It can be proven that such an air hole has an approximate fundamental ($HE_{11}$) modal index of $\sim \sqrt{1-(u_{01}\lambda/\pi d')^2}$, being the same with that of an air slab with the thickness of $d'\pi/2u_{01} \approx d_{opt}$. We evenly arrange eight nearly-touched glass tubes surrounding the core of the NCF. The glass wall thicknesses of both structures are 0.24μm, and the inter-tube gaps in the NCF are 0.8μm wide. This is close to a practically realizable NCF with current state-of-art fabrication technique [3-7]. For the NCF, in order to retrieve the complex effective indices, $n_{eff}$, of the fundamental core modes, we perform numerical simulation based on a finite-element mode solver (Comsol Multiphysics). The mesh size and the perfectly matched layer are optimized [48] to ensure calculation accuracy.

In Fig. 3, we compare the attenuation spectra of the two types of fibers layer by layer. For N = 1, both fiber structures exhibit smooth spectra with no resonance at very similar level (350-400dB/m at 800nm, black curves). It is manifest that such a level of light leakage is from the (Fresnel) transmission across one interface at near-grazing incidence. For N = 2, by adding one glass layer, the shapes of the attenuation spectra alter and show a fundamental transmission band starting from ~520nm (blue curves). This is a result of the wavelength-dependent ARROW effect in the glass layer. The loss figure decreases to 13.8dB/m at 800nm for the annular fiber (unless specified, the wavelength will be fixed to 800nm hereafter). From above analysis, we understand that this ~14dB loss figure reduction consists of ~8dB from the addition of one interface and ~6dB from the addition of one glass layer with nearly optimum ARROW condition. However, surprisingly, for the NCF the loss figure drops to 2.1dB/m, a dramatic reduction of 22.8dB. This phenomenon indicates that, apart from the three effects identified in the last section, the shape of glass wall may also play an important role in light guidance. The physical origin of this new effect will be discussed in the next paragraph. Next, from N = 2 to N = 3, an anti-resonant reflection layer of air is added for both fibers. In contrast to the glass layer, the effect of this air layer is wavelength insensitive and leads to a downward shift of attenuation spectrum from the blue curves to the green curves. Although the shapes of the air layer are completely different, the conditions of $U_{air} \approx$ 0.5 are both fulfilled and the loss figures are identically reduced by 14.4dB and 14.3dB to 0.5dB/m and 0.078dB/m respectively. In comparison to the glass layer, the shape of air layer seems not sensitive to the attenuation, which will also be analyzed in the next paragraph. In the last step, from N = 3 to N = 4, a second glass layer is added. It contributes another 13.6dB and 17.5dB reduction to the loss figures for the annular fiber and the NCF respectively, reaching to the values of 0.022dB/m and 0.0014dB/m (yellow curves). Again, the influence of the shape of glass wall appears. The result hints that besides the negatively curved core surround, which has been widely recognized as the origin of the superiority of tube lattice ARF [26], a positively curved glass wall in the cladding also can suppress light leakage. The shape of glass wall, therefore, could be the 4$^{th}$ engineering degree of freedom for CL in ARF.

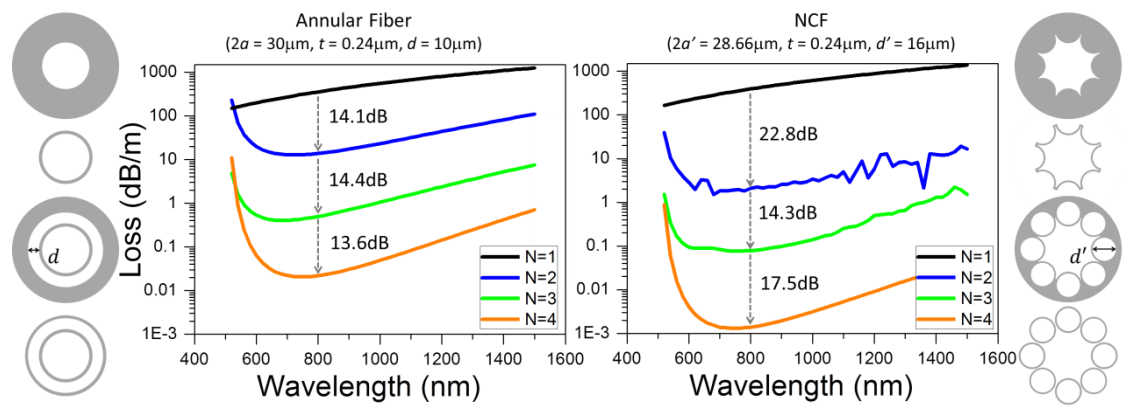

**Fig. 3**. Calculated and simulated attenuation spectra of annular fiber and NCF with the interface numbers N = 1, 2, 3, 4 respectively. The geometrical parameters are list. For the NCF, the core diameter is measured as the maximum inscribed circle. The variations of loss figure are listed at 800nm. For the spectra of NCF, the blue curve shows fast undulations. This is caused by the Fano resonances introduced at the breaking ends because of our disassembly of the glass tube.

To elucidate the aforementioned phenomenon, we resort to the energy conservation principle. In a lossless dielectric fiber, the attenuated light energy in the longitudinal direction is equal to the light leakages in the radial directions. By integrating all the outwardly-propagating beams in the transverse plane, fiber's attenuation coefficient can be inferred. In our previous work [40, 41], we have pointed out that if one chooses an equiphase closed loop in the outermost medium as the path of integration and acquires the electromagnetic fields on it, the radiation intensities in all the radial directions could be derived by utilizing Green's theorem [49]. Fig. 4(a) shows that the outermost boundary of a glass wall indeed has the effect of dragging phase contours [50]. Therefore, we can qualitatively interpret the reduction (modification) of the attenuation coefficients for a curved ARF compared with an annular fiber. The phases of the optical rays escaping from the core are noticeably dragged by the glass wall, which suppresses the sequent radiations to the far field by constituting some destructive interference. On the contrary, a circular eqiphase contour formed in the annular fiber is relatively favored for outward radiations or leakages. In Fig. 4(b), the attenuation spectra of three single-wall ARFs (the annular fiber, the negatively-curved and the positively-curved fibers) are compared at the prerequisite of identical modal indices. A difference of 8-11dB at 800nm is manifest, verifying the engineering capability of the shape of glass wall for light confinement. Again here we see that negative curvature is very likely not the best option for CL reduction. Notice that in the transverse plane and in the outermost medium of homogenous air, the electromagnetic field is governed by a two-dimensional Helmholtz equation $\nabla_T^2 \mathbf{E} + k_T^2 \mathbf{E} = 0$ with $k_t \approx k_0 \sqrt{1 - \mathrm{Re}(n_{eff})^2} \approx u_{01}/a$ and $a$ the core radius. The dimension of the closed loop of integration is of the same magnitude as $a$. Therefore, the phase retardation accumulated in the integration loop has possibility to construct destructive interference if the shape of the outermost boundary of glass wall is properly designed. The same situation occurs in the case of Fig. 4(c) with two layers of glass wall involved. Here, the fact that the dimension of the outermost boundary is prominently bigger than the core size makes the extent of the destructive interference modest, with a reduction of attenuation coefficient of ~3dB. On the other hand, if the outermost medium changes to silica, i.e., for the case of N = 1,3…, the transverse wavevector in the Helmholtz equation becomes $K_t = k_0 \sqrt{n^2 - \mathrm{Re}(n_{eff})^2} \gg k_t$. The phase retardation across the integration loop will be magnitude larger and will wipe away any interferometric effect, which is the situation of the black and the green spectra in Fig. 3.

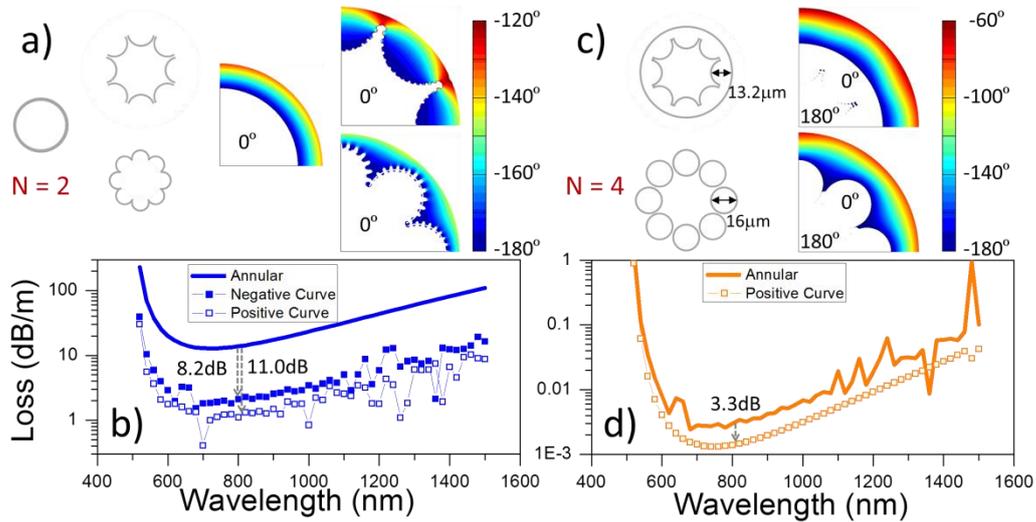

**Fig. 4**. (a,c) Simulated phase contours of the electric field for different fiber structures and (b,d) their effects on CL. In (a), the color bar is from -120° to -180°. The phases inside the core regions are uniformly 0° and are left in white. One can see that the phase contours of -180° (in dark blue and at the edges of the white color regions) exactly follow the outermost boundaries of glass walls. (c) is plotted in similar way. In (b) and (d), the simulated attenuation spectra are given and all these structures have identical modal indices (not shown).

So far, from N = 1 to N = 4, all the glass and air 'layers' are designed near the optimum ARROW condition $U_{glass}$ = 0.5 (for the central wavelength region) and $U_{air}$ = 0.5. When we consider a realistic structure, i.e. N = 5 with the outermost structure to be a jacketing tube, the degree of freedom for tailoring the thickness of the outermost air layer is restricted. Since the jacketing tube has to contact with the glass tubes (Fig. 5(d)), a layer of air with no ARROW benefit is incorporated. Based on Fig. 2, we probably can evaluate the detrimental influence of adding this air layer. In Fig. 5, we adjust the air layer thicknesses in the annular fiber and the NCF by altering $d_2$ and $g$ respectively. The loss figures in the case of N = 4 (circle symbols) are taken as a reference. When $d_2$ and $g$ approach 10μm, the antiresonant reflection condition $U_{air}$ = 0.5 is fulfilled and the loss figures are dropped by 14.1dB and 13.8dB respectively. However, for small values of $d_2$ and $g$, $U_{air}$ is far from the optimum value and the multi-path interference in the air layer results in an enhancement of light leakage as indicated in Fig.2. For the NCF, when $g$ = 0, the loss figure, compared to the case of N = 4, increases by 6.4dB (black curve in Fig. 5(f)), similar to the effect of $d_2$ = 0.45μm ($U_{air}$=0.023) in the annular fiber (black curve in Fig. 5(e)). The situation becomes worse when we consider the fact that the glass tubes in NCF have to merge into the outer jacket (see Fig. 5(d)), which is the common case in practical fabrication. For $g$ = -0.24μm, the loss figure of NCF is increased by 8.1dB (red curve in Fig. 5(f)), corresponding to $d_2$ = 0.3μm ($U_{air}$ = 0.015) in annular fiber (red curve in Fig. 5(e)). In this practical case, the minimum simulated CL of the single-ring NCF becomes 7.5dB/km at 700nm. The reported fabricated fibers show similar loss levels [3-7]. Based on above analysis, how to prevent the outermost air layer to be in the resonant leakage region will be a crucial challenge [47]. Note this resonant light leakage is very sensitive to the air layer thickness but insensitive to the wavelength. With the adoption of our model, we hope more complicated phenomenon about ARFs can be explained. In many recent designs, more inter-medium interfaces and irregular glass wall shapes have been introduced for achieving a lower loss level, such as the nested structure in [20, 51] and the elliptical tube structure in [52, 53].

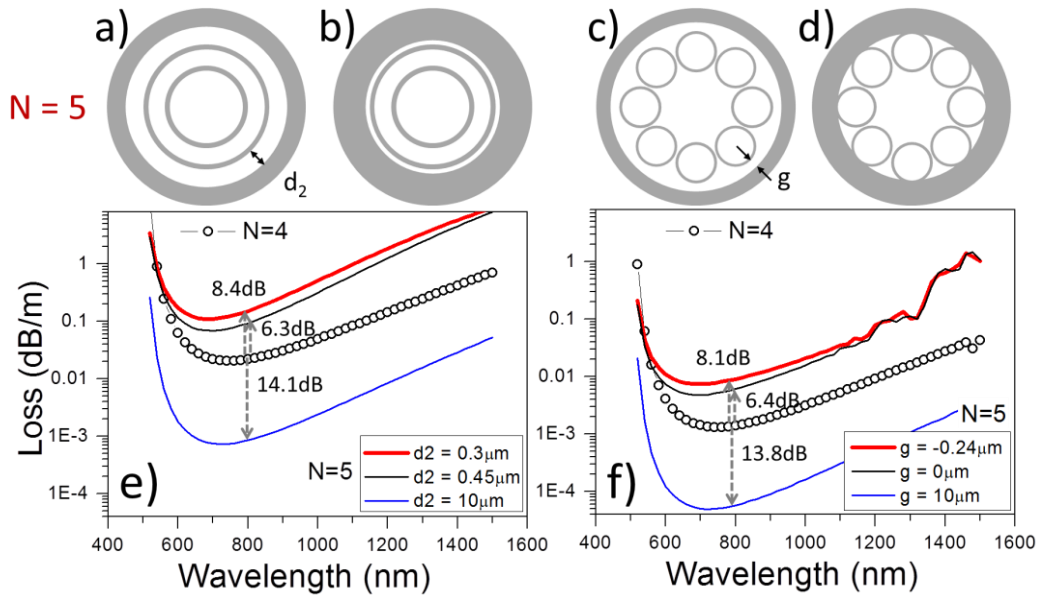

**Fig. 5**. Comparison of annular fiber and NCF with the interface number N = 5. (a) and (c) denote the parameters $d_2$ and $g$. (b) and (d) respectively show the cases of $d_2$ = 0.3μm for annular fiber

and $g$ = -0.24μm for practical NCF when the 8 tubes merge into the jacket. (e) and (f) plot their attenuation spectra for varying $d_2$ and $g$.

## 4. Discussion and conclusion

To elucidate the light guidance in NCF, we comply with the cylindrical symmetry and compare with the multi-layered annular fiber by disassembling the NCF structure into two glass layers and two air layers. A multi-layered model is built where the origins of light confinement are decomposed into four effects: single-path Fresnel transmission through cascaded interfaces, near-grazing incidence, multi-path interference caused by Fresnel reflection (ARROW), and glass wall shape. The first two effects can be analytically expressed as an exponential function of the number of interfaces and are relevant to the refractive index contrast. In the case of small glancing angle, small amount of Fresnel transmission ensures the general tone of effective light confinement. We have yet dug in, but the polarization dependence of the Fresnel transmittance, which is expressed in Eq. (1) as a function of the refractive index contrast, obviously deserves further studies for vector mode selection function [43]. The third and fourth effects play the role of further enhancing or harnessing light confinement depending on different interference conditions. The third effect, i.e. the ARROW effect, constructs interference in the directions perpendicular to the layers. While the fourth effect, i.e. the one relevant to the glass wall shape, builds interference in the transverse plane along a closed loop near the outermost boundary of glass wall. Based on numerical calculation, we point out that the FOM achievable by the ARROW effect is only ~6dB for each layer, often being overestimated. The virtue of the ARROW model is actually to direct to the suppression of resonant light leakages. On the contrary, the shape of glass wall can serve as a new engineering paradigm, giving rise to considerable benefits for reducing CL (8-11dB in the case of Fig. 4). Moreover，the ARROW FOM diagram itself, like the one shown in Fig. 2, also deserves further investigations, e.g. in the higher order bands.

In conclusion, concerted operations of four effects give rise to a low confinement loss in single-ring nodeless NCF. Our multi-layered analytic model can not only provide an insightful picture for understanding the behavior of this attractive fiber structure but also offer a useful tool to design new functions in this platform.

## Appendix A: Derivation of Equation (1)

To derive Eq. (1), we use Snell's law and Fresnel formula

$$n_1 \sin i_1 = n_2 \sin i_2, \quad \begin{cases} \dfrac{I_s^{(r)}}{I_s} = \left|\dfrac{\sin(i_1 - i_2)}{\sin(i_1 + i_2)}\right|^2 \\ \dfrac{I_p^{(r)}}{I_p} = \left|\dfrac{\tan(i_1 - i_2)}{\tan(i_1 + i_2)}\right|^2 \end{cases}, \quad \begin{cases} \dfrac{I_s^{(t)}}{I_s} = 1 - \dfrac{I_s^{(r)}}{I_s} \\ \dfrac{I_p^{(t)}}{I_p} = 1 - \dfrac{I_p^{(r)}}{I_p} \end{cases}. \quad (4)$$

Here the incidence and the refraction angles $i_{1,2}$ are relative to the normal direction of the interface, complementary to the glancing angle $\theta_z/\Theta_z$ (see Fig. 1). The subscripts ($s/p$) stand for the two polarizations, and the superscripts ($r/t$) represent the reflection and transmission. When we reverse the direction of the optical ray, i.e. $i_1 \Leftrightarrow i_2$, $n_1 \Leftrightarrow n_2$, the results of above equation do not change, which means every interface in Fig. 1 has the same transmittance regardless of the incidence direction from glass to air or from air to glass. Assuming the glancing angle in the air layer $\theta_z \ll 1$, to the leading order terms, we can derive $\frac{I_s^{(t)}}{I_s} \approx \frac{4 \cdot \sin\theta_z}{\sqrt{n^2-1}}$ and $\frac{I_p^{(t)}}{I_p} \approx 4 \cdot \sin\theta_z \cdot \left(\frac{1}{\sqrt{n^2-1}} + \sqrt{n^2-1}\right)$, i.e. Eq. (1).

## Appendix B: Derivation of the maximum FOM of ~6dB

In a recently-published paper [42], an analytic expression is derived for the CL of the annular fiber. By utilizing the asymptotic form of Hankel functions, the author gives the following expression for the attenuation coefficient of the fundamental $HE_{11}$ mode,

$$\alpha[dB/m] \approx \frac{10}{a \cdot \ln 10}\left(\frac{u_{01}\lambda}{2\pi a}\right)^{N+1}\left\{\left(\frac{1}{\sqrt{n^2-1}}\right)^N + \left(\frac{n^2}{\sqrt{n^2-1}}\right)^N\right\}\prod_{i=1}^{N-1}\frac{1}{\sin^2\phi_i}. \quad (5)$$

In that paper, $\phi_i^{(g)}$ and $\phi_i^{(a)}$ are respectively defined to be $2\pi t_i\sqrt{n^2-1}/\lambda$ and $u_{01}d_i/a$, and the ARROW effect is embodied in $\prod_{i=1}^{N-1}\frac{1}{\sin^2\phi_i}$. While, it is not difficult to prove that $\phi_i^{(g,a)} \approx \pi \cdot U_{glass,air}$. Comparing Eq. (5) with Eq. (2) and noticing that $\sin\theta_z \approx \sqrt{1-\mathrm{Re}(n_{eff})^2} \approx u_{01}\lambda/2\pi a$ yield that the maximum FOM of the ARROW effect from one dielectric layer is the factor 4 or ~6dB!

## Acknowledgments

This work was supported by the National Key R&D Program of China (No. 2017YFA0303800), the National Natural Science Foundation of China (No. 61575218, 61377098, 61675011, 61527822, 61535009), and the Instrument Developing Project of the Chinese Academy of Sciences (No. YZ201346).

## References

1. R. F. Cregan, B. J. Mangan, J. C. Knight, T. A. Birks, P. St.J. Russell, P. J. Roberts and D. C. Allan, "Single mode photonic band gap guidance of light in air," Science 285, 1537-1539 (1999).

2. P. Russel, "Photonic Crystal Fibers", Science 299, 358-362 (2003).

3. B. Debord, A. Amsanpally, M. Chafer, A. Baz, M. Maurel, J. M. Blondy, E. Hugonnot, F. Scol, L. Vincetti, F. Gérôme, and F. Benabid, "Ultralow transmission loss in inhibited-coupling guiding hollow fibers," Optica 4, 209–217 (2017).

4. P. Uebel, M. C. Günendi, M. H. Frosz, G. Ahmed, N. N. Edavalath, J.-M. Ménard, and P. St.J. Russell, "Broadband robustly single-mode hollow-core PCF by resonant filtering of higher-order modes," Opt. Lett. 41, 1961–1964 (2016).

5. J. R. Hayes, S. R. Sandoghchi, T. D. Bradley, Z. Liu, R. Slavik, M. A. Gouveia, N. V. Wheeler, G. Jasion, Y. Chen, E. N. Fokoua, M. N. Petrovich, D. J. Richardson, and F. Poletti, "Antiresonant hollow core fiber with an octave spanning bandwidth for short haul data communications," J. Lightwave Technol. 35, 437–442 (2017).

6. M. Michieletto, J. K. Lyngs, C. Jakobsen, J. Lgsgaard, O. Bang, and T. T. Alkeskjold, "Hollow-core fibers for high power pulse delivery," Opt. Express 24, 7103–7119 (2016).

7. S. Gao, Y. Y. Wang, X. Liu, C. Hong, S. Gu, and P.Wang, "Nodeless hollow-core fiber for the visible spectral range," Opt. Lett. 42, 61–64 (2017).

8. P. Jaworski, F. Yu, R. R. J.Maier,W. J.Wadsworth, J. C. Knight, J. D. Shephard, and D. P. Hand, "Picosecond and nanosecond pulse delivery through a hollow-core negative curvature fiber for micro-machining applications," Opt. Express 21, 22742–22753 (2013).

9. Y. Wang, M. Alharbi, T. D. Bradley, C. Fourcade-Dutin, B. Debord, B. Beaudou, F. Gérôme, and F. Benabid, "Hollow-core photonic crystal fibre for high power laser beam delivery," High Power Laser Sci. Eng. 1, 17–28 (2013).

10. U. Elu, M. Baudisch, H. Pires, F. Tani, M. H. Frosz, F. Köttig, A. Ermolov, P. St.J. Russell, and J. Biegert, "High average power and single-cycle pulses from a mid-IR optical parametric chirped pulse amplifier," Optica 4, 1024-1029 (2017).

11. T. Balciunas, C. Fourcade-Dutin, G. Fan, T. Witting, A. A. Voronin, A. M. Zheltikov, F. Gérôme, G. G. Paulus, A. Baltuska, and F. Benabid, "A strong-field driver in the single-cycle regime based on self-compression in a kagome fibre," Nat. Commun. 6, 6117 (2015).

12. Felix Köttig, Francesco Tani, Christian Martens Biersach, John C. Travers, and Philip St.J. Russell, "Generation of microjoule pulses in the deep ultraviolet at megahertz repetition rates," Optica 4, 1272-1276 (2017).

13. F. Amrani, F. Delahaye, B. Debord, L. L. Alves, F. Gerome, and F. Benabid, "Gas mixture for deep-UV plasma emission in a hollow-core photonic crystal fiber," Opt. Lett. 42, 3363-3366 (2017)

14. M. R. A. Hassan, F. Yu, W. J. Wadsworth, and J. C. Knight, "Cavity-based mid-IR fiber gas laser pumped by a diode laser," Optica 3, 218–221 (2016).

15. A. M. Cubillas, X. Jiang, T. G. Euser, N. Taccardi, B. J. M. Etzold, P. Wasserscheid, and P. St.J. Russell, "Photochemistry in a soft-glass single-ring hollow-core photonic crystal fiber," Analyst 142, 925–929 (2017).

16. X. Liu, W. Ding, Y. Y. Wang, S. Gao, L. Cao, X. Feng, and P. Wang, "Characterization of a liquid-filled nodeless anti-resonant fiber for biochemical sensing," Opt. Lett. 42, 863–866 (2017).

17. S. Okaba, T. Takano, F. Benabid, T. Bradley, L. Vincetti, Z. Maizelis, V. Yampol'skii, F. Nori, and H. Katori, "Lamb-Dicke spectroscopy of atoms in a hollow-core photonic crystal fibre," Nat. Commun. 5, 4096 (2014).

18. R. R. Gattass, D. Rhonehouse, D. Gibson, C. C. McClain, R. Thapa, V. Q. Nguyen, S. S. Bayya, R. J. Weiblen, C. R. Menyuk, L. B. Shaw, and J. S. Sanghera, "Infrared glass-based negative-curvature anti-resonant fibers fabricated through extrusion," Opt. Express 24, 25697–25703 (2016).

19. J. Yang, J. Zhao, C. Gong, H. Tian, L. Sun, P. Chen, L. Lin, and W. Liu, "3D printed low-loss THz waveguide based on Kagome photonic crystal structure," Opt. Express 24, 22454–22460 (2016).

20. F. Poletti, "Nested antiresonant nodeless hollow core fiber," Opt. Express 22, 23807–23828 (2014).

21. T. A. Birks, P. J. Roberts, P. S. J. Russell, D. M. Atkin, and T. J. Shepherd, "Full 2-d photonic bandgaps in silica/air structures," Electron. Lett., vol. 31, no. 22, pp. 1941–1943, 1995.

22. P. Roberts, F. Couny, H. Sabert, B. Mangan, D. Williams, L. Farr, M. Mason, A. Tomlinson, T. Birks, J. Knight, and P. St.J. Russell, "Ultimate low loss of hollow-core photonic crystal fibres," Opt. Express 13, 236–244 (2005).

23. F. Benabid, J. C. Knight, G. Antonopoulos, and P. St.J. Russell, "Stimulated Raman scattering in hydrogen-filled hollow-core photonic crystal fiber," Science 298, 399–402 (2002).

24. F. Benabid and P. J. Roberts, "Linear and nonlinear optical properties of hollow core photonic crystal fiber," J. Mod. Opt. 58, 87–124 (2011).

25. Y. Wang, F. Couny, P. J. Roberts, and F. Benabid, "Low loss broadband transmission in optimized core-shape Kagome hollow-core PCF," in Conference on Lasers and Electro-Optics (CLEO, 2010), paper CPDB4.

26. Y. Y. Wang, N. V. Wheeler, F. Couny, P. J. Roberts, and F. Benabid, "Low loss broadband transmission in hypocycloid-core Kagome hollow-core photonic crystal fiber," Opt. Lett. 36, 669–671 (2011).

27. Y. Y. Wang, X. Peng, M. Alharbi, C. F. Dutin, T. D. Bradley, F. Gérôme, M. Mielke, T. Booth, and F. Benabid, "Design and fabrication of hollow-core photonic crystal fibers for high-power ultrashort pulse transportation and pulse compression," Opt. Lett. 37, 3111–3113 (2012).

28. N. V. Wheeler, T. D. Bradley, J. R. Hayes, M. A. Gouveia, S. Liang, Y. Chen, S. R. Sandoghchi, S. M. Abokhamis Mousavi, F. Poletti, M. N. Petrovich, and D. J. Richardson, "Low-loss Kagome hollow-core fibers operating from the near to the mid-IR," Opt. Lett. 42, 2571–2574 (2017).

29. A. D. Pryamikov, A. S. Biriukov, A. F. Kosolapov, V. G. Plotnichenko, S. L. Semjonov, and E. M. Dianov, "Demonstration of a waveguide regime for a silica hollow-core microstructured optical fiber with a negative curvature of the core boundary in the spectral region >3.5 μm," Opt. Express 19, 1441–1448 (2011).

30. F. Yu, W. J.Wadsworth, and J. C. Knight, "Low loss silica hollow core fibers for 3–4 μm spectral region," Opt. Express 20, 11153–11158 (2012).

31. A. N. Kolyadin, A. F. Kosolapov, A. D. Pryamikov, A. S. Biriukov, V. G. Plotnichenko, and E. M. Dianov, "Light transmission in negative curvature hollow core fiber in extremely high material loss region," Opt. Express 21, 9514–9519 (2013).

32. A. F. Kosolapov, G. K. Alagashev, A. N. Kolyadin, A. D. Pryamikov, A. S. Biryukov, I. A. Bufetov, and E. M. Dianov, "Hollow-core revolver fibre with a double-capillary reflective cladding," Quantum Electron. 46, 267–270 (2016).

33. A. Hartung, J. Kobelke, A. Schwuchow, K. Wondraczek, J. Bierlich, J. Popp, T. Frosch, and M. A. Schmidt, "Double antiresonant hollow core fiber-guidance in the deep ultraviolet by modified tunneling leaky modes," Opt. Express 22, 19131–19140 (2014).

34. E. Marcatili and R. Schmeltzer, "Hollow metallic and dielectric waveguides for long distance optical transmission and lasers," Bell Syst. Tech. J., vol. 43, no. 4, pp. 1783–1809, (1964).

35. P. Yeh, A. Yariv, and E. Marom, "Theory of Bragg fiber," J. Opt. Soc. Am. 68, 1196–1201 (1978).

36. N. M. Litchinitser, A. K. Abeeluck, C. Headley, and B. J. Eggleton, "Antiresonant reflecting photonic crystal optical waveguides," Opt. Lett. 27, 1592–1594 (2002).

37. F. Couny, F. Benabid, P. J. Roberts, P. S. Light, and M. G. Raymer, "Generation and photonic guidance of multi-octave optical-frequency combs," Science 318, 1118–1121 (2007).

38. L. Vincetti and V. Setti, "Waveguiding mechanism in tube lattice fibers," Opt. Express 18, 23133–23146 (2010).

39. G. J. Pearce, G. S. Wiederhecker, C. G. Poulton, S. Burger, and P. St.J. Russell, "Models for guidance in kagome-structured hollow-core photonic crystal fibres," Opt. Express 15, 12680–12685 (2007).

40. W. Ding and Y. Wang, "Analytic model for light guidance in single wall hollow-core anti-resonant fibers," Opt. Express 22, 27242–27256 (2014).

41. W. Ding and Y. Wang, "Semi-analytical model for hollow-core anti-resonant fibers," Front. Phys. 3, 16 (2015).

42. D. Bird, "Attenuation of model hollow-core, anti-resonant fibres," Opt. Express 25, 23215-23237 (2017).

43. M. Zeisberger and M. A. Schmidt, "Analytic model for the complex effective index of the leaky modes of tube-type anti-resonant hollow core fibers", Sci. Rep. 7, 11761 (2017).

44. S. Gao, Y. Wang, X. Liu, W. Ding, and P. Wang, "Bending loss characterization in nodeless hollow-core anti-resonant fiber," Opt. Express 24, 14801–14811 (2016).

45. D. Marcuse, Theory of Dielectric Optical Waveguides, 2nd ed. (Academic, 1991).

46. J. Hu and C. R. Menyuk, "Understanding leaky modes: slab waveguide revisited," Adv. Opt. Photon. 1, 58–106 (2009).

47. J. R. Hayes, F. Poletti, M. S. Abokhamis, N. V. Wheeler, N. K. Baddela, and D. J. Richardson, "Anti-resonant hexagram hollow core fibers," Opt. Express 23, 1289–1299 (2015).

48. S. Selleri, L. Vincetti, A. Cucinotta, and M. Zoboli, "Complex FEM modal solver of optical waveguides with PML boundary conditions," Opt. Quantum Electron. 33(4-5), 359–371 (2001).

49. M. Born and E. Wolf, Principles of Optics: Electromagnetic Theory of Propagation, Interference and Diffraction of Light, 6th ed. (Cambridge University Press, 1999).

50. W. Ding and Y. Y. Wang, "Hybrid transmission bands and large birefringence in hollow-core anti-resonant fibers," Opt. Express 23(16), 21165–21174 (2015).

51. M. I. Hasan, N. Akhmediev, and W. Chang, "Positive and negative curvatures nested in an antiresonant hollow-core fiber," Opt. Lett. 42, 703–706 (2017).

52. M. S. Habib, O. Bang, and M. Bache, "Low-loss single-mode hollow-core fiber with anisotropic anti-resonant elements," Opt. Express 24, 8429–8436 (2016).

53. F. Meng, B. Liu, Y. Li, C.Wang, and M. Hu, "Low loss hollow-core antiresonant fiber with nested elliptical cladding elements," IEEE Photon. J. 9, 7100211 (2017).